\documentclass[%
 aip,
 amsmath,amssymb,
 reprint,%
]{revtex4-1}

\usepackage{graphicx}
\usepackage{dcolumn}
\usepackage{bm}

\usepackage[utf8]{inputenc}
\usepackage[T1]{fontenc}
\usepackage{mathptmx}
\usepackage{etoolbox}

\usepackage[mathscr]{eucal}

\usepackage[dvipsnames]{xcolor}

\usepackage{xcolor}
\usepackage{lineno}

\usepackage{amsmath,amssymb}
\usepackage[utf8]{inputenc}
\usepackage[T1]{fontenc}
\usepackage{mathptmx}

\usepackage{sidecap}

\usepackage{graphicx,scalerel}

\makeatletter
\def\@email#1#2{%
 \endgroup
 \patchcmd{\titleblock@produce}
  {\frontmatter@RRAPformat}
  {\frontmatter@RRAPformat{\produce@RRAP{*#1\href{mailto:#2}{#2}}}\frontmatter@RRAPformat}
  {}{}
}%
\makeatother
\begin{document}

\preprint{AIP/123-QED}

\title[]{Numerical estimate of the viscous damping of capillary-gravity waves: A macroscopic depth-dependent slip-length model}

\author{A. Bongarzone}
\author{F. Gallaire}%
\email{francois.gallaire@epfl.ch.}
\affiliation{ 
Laboratory of Fluid Mechanics and Instabilities, École Polytechnique Fédérale de Lausanne, Lausanne, CH-1015, Switzerland
}%

\date{\today}

\begin{abstract}
We propose a numerical approach to regularize the contact line singularity appearing in the computation of viscous capillary-gravity waves with moving contact line in cylindrical containers.  The linearized Navier-Stokes equations are complemented by a macroscopic Navier-like slip condition on the container side wall, with a depth-varying slip-length bridging a free-slip condition at the contact line to a no-slip condition further away from the meniscus. In accordance with suggestions from the literature, this characteristic penetration depth is chosen as the typical Stokes layer thickness pertaining to the eigenfrequency. Since the latter is unknown, the resulting nonlinear eigenvalue problem is first simplified using the inviscid eigenfrequency to estimate the  Stokes layer thickness. The solution is then shown to provide consistent results when compared to the asymptotic approaches of the literature using the inviscid eigenmodes to estimate the different sources of dissipation with the exception of the neglected corner regions. This demonstrates that such numerical regularization scheme is suitable to retrieve physically meaningful results.
\end{abstract}

\maketitle

\section{\label{sec:Sec1}Introduction}

Sloshing motion represents an archetypal damped oscillator in fluid mechanics. The eigenfrequencies of capillary-gravity waves in closed basins, e.g. in circular cylinder, were traditionally derived in the potential flow limit \cite{Lamb32}, while the linear viscous dissipation at the free-surface, at the solid walls and in the bulk for low-viscosity fluids was typically accounted for by a boundary layer approximation \cite{Case1957,Miles67}. These classic theoretical approaches are built on the assumption that the free liquid surface, $\eta$, intersects the lateral wall orthogonally, and the contact line can freely slip with constant zero slope, 
\begin{equation}
\label{eq:FCL0}
\partial_n\eta=0\ \ \ \ \text{(free-end edge condition)},
\end{equation}
\noindent where $\partial_n$ is the spatial derivative in the direction normal to the lateral wall.\\
\indent However, when the fluid viscosity is formally accounted for, e.g. in a numerical scheme, the boundary conditions on the tangential velocity components at the lateral solid wall require a special treatment, in order to avoid the contact line singularity.\\
\indent With regard to circular cylinders, at the sidewall, $r=R$ ($r$ is the radial coordinate and $R$ is the container radius), a slip-length model is typically adopted, thus assuming that the fluid speed relative to the solid wall is proportional to the viscous stress \cite{navier1823memoire,Lauga2007,Viola2018b}
\begin{equation}
\label{eq:SLclassicPhiZ}
{u}_{\phi}+l_s\left(\partial_r{u}_{\phi}-{u}_{\phi}/r\right)=0,\ \ \ {u}_{z}+l_s\partial_r{u}_z=0,
\end{equation}
\noindent with $u_{\phi}$ and $u_z$ the tangential azimuthal and axial velocity components, respectively. Such conditions are indeed required in order to regularize the stress-singularity arising in the viscous solution of a moving contact line over a solid substrate, for which the no-slip condition is incompatible \cite{Huh71,Davis1974}. In its original formulation, the slip-length $l_s$ appearing in~\eqref{eq:SLclassicPhiZ} is assumed finite and constant in space along the whole sidewall. The numerical implementation of conditions~~\eqref{eq:FCL0}-\eqref{eq:SLclassicPhiZ} with constant $l_s$ has been proposed Viola \& Gallaire (2018) \cite{Viola2018b}, where a non-dimensional value of $l_s=5\times10^{-4}$, corresponding to a dimensional value of $l_sR=25\,\text{$\mu$m}$ for a container radius $R=0.05\,\text{m}$, was used. The numerically estimated damping coefficient associated with the first non-axisymmetric mode was then compared with the analytical prediction by Case \& Parkinson (1957) \cite{Case1957} (see Sec.~\ref{sec:Sec5}), who assume a perfect no-slip condition at the wall (the stress-singularity at the contact line is simply ignored in their asymptotic analysis). However, the numerical damping strongly underestimates the one predicted according to Case \& Parkinson (1957)\cite{Case1957}, which instead agrees fairly well with their experimental measurements. The origin of such a disagreement can be tentatively ascribed to the fact that a slip-length value of $25\,\text{$\mu$m}$ is significantly larger than the one observed experimentally \cite{thompson1989simulations}, i.e. $0.1-10\,\text{nm}$. Lower values of $l_s$ could not be employed in Ref.~\onlinecite{Viola2018b} due to the grid resolution needed to guarantee numerical convergence of the results.\\
\indent Nevertheless, there is actually a more subtle aspect intrinsic to the formulation and the numerical implementation of conditions~\eqref{eq:FCL0}-\eqref{eq:SLclassicPhiZ}. The dynamics of small-amplitude viscous capillary-gravity waves is essentially governed by the unsteady Stokes equations (see Sec.~\ref{sec:Sec2}), which describe the problem from a macroscopic perspective, where the microscopic (molecular) dynamics at play at the contact line is completely filtered out. The free-end edge contact line condition has indeed only meaning in a macroscopic framework. It follows that the slip-length $l_s$ must also assume the significance of a phenomenological macroscopic slip-length, which only serves to regularize the contact line stress-singularity and which is not intrinsically linked to the actual molecular slip-length. When conditions~\eqref{eq:FCL0} and~\eqref{eq:SLclassicPhiZ} are simultaneously imposed in a numerical scheme, a paradox concerning the tangential velocity components arises.\\
\indent A straightforward way to highlight such a paradox consists in combining the free-end edge condition~\eqref{eq:FCL0}, $\partial_r\eta=0$, with the linearized kinematic equation, $\partial_t\eta=u_z$, differentiated with the respect to the radial direction, 
\begin{equation}
\label{eq:KINdr0}
\partial_t \left(\partial_r{\eta}\right)=\partial_r {u}_z\ \ \ \rightarrow\ \ \  \partial_r {u}_z =0,
\end{equation}
\noindent Relation~\eqref{eq:KINdr0} points out that, by approaching the contact line from the sidewall, one must impose a sidewall stress-free condition on the axial velocity component ${u}_z$.\\
\indent A generalization to cylindrical coordinates of the more rigorous reasoning, based on the vorticity argument, outlined by Miles (1990) \cite{miles1990capillary}, suggests indeed that, at the contact line, the only sidewall boundary condition compatible with the free-end edge condition is a stress-free condition for both tangential velocity components, ${u}_z$ and ${u}_{\phi}$.\\
\indent In contradistinction with the argument above, the attempt to impose a constant and, at the same time, very small (experimentally inspired) slip-length $l_s$ along the entire wall, will translate in nearly no-slip condition everywhere at the sidewall,
\begin{equation}
\label{eq:consLS}
{u}_z+\underbrace{l_s}_{\ll1}{u}_z=0\ \ \ \longrightarrow\ \ \ {u}_z\approx0,
\end{equation}
\noindent which is not compatible with the macroscopic free-end contact line condition and which may therefore lead to a numerical unstable solution or to incorrect estimates of the damping coefficients. Obviously, a stress-free condition everywhere would represent a mathematically correct alternative, but its imposition would automatically filter out the corresponding viscous boundary layer, leading to a strong numerical underestimation of the viscous damping \cite{Case1957}.\\
\indent With the aim to marry the simultaneous need for a sufficiently accurate description of the viscous sidewall boundary layer and a formally correct treatment of the contact line region, we propose in this work the adoption of a phenomenological macroscopic and depth-dependent slip-length model, which constitutes a practical, \textit{ad}--\textit{hoc}, boundary condition for numerical calculations.\\
\indent A similar slip-length model was proposed by Ting \& Perlin (1995) \cite{ting1995boundary} (see their Sec.~4.6 and~5). Such a model was developed on the base of their experimental observations for a vertically oscillating plate. In their model, wall-slip occurs within a specific distance from the contact line, whereas the flow obeys the no-slip condition outside this slip region.\\
\indent The manuscript is organized as follows. The flow configuration and the physical model governing the problem are given in Sec.~\ref{sec:Sec2}. The adopted depth-varying slip-length model, core of the present work, is introduced in Sec.~\ref{sec:Sec3}. Sec.~\ref{sec:Sec4} is devoted to the description of the numerical method on which results commented in Sec.~\ref{sec:Sec5} are based on. Sensitivity to the free parameters of the slip-length model is explored in Sec.~\ref{sec:Sec6}. Lastly, final conclusions and comments are outlined in Sec.~\ref{sec:Sec7}.

\section{\label{sec:Sec2}Configuration and governing equations}

\begin{figure}
\includegraphics[scale=0.3]{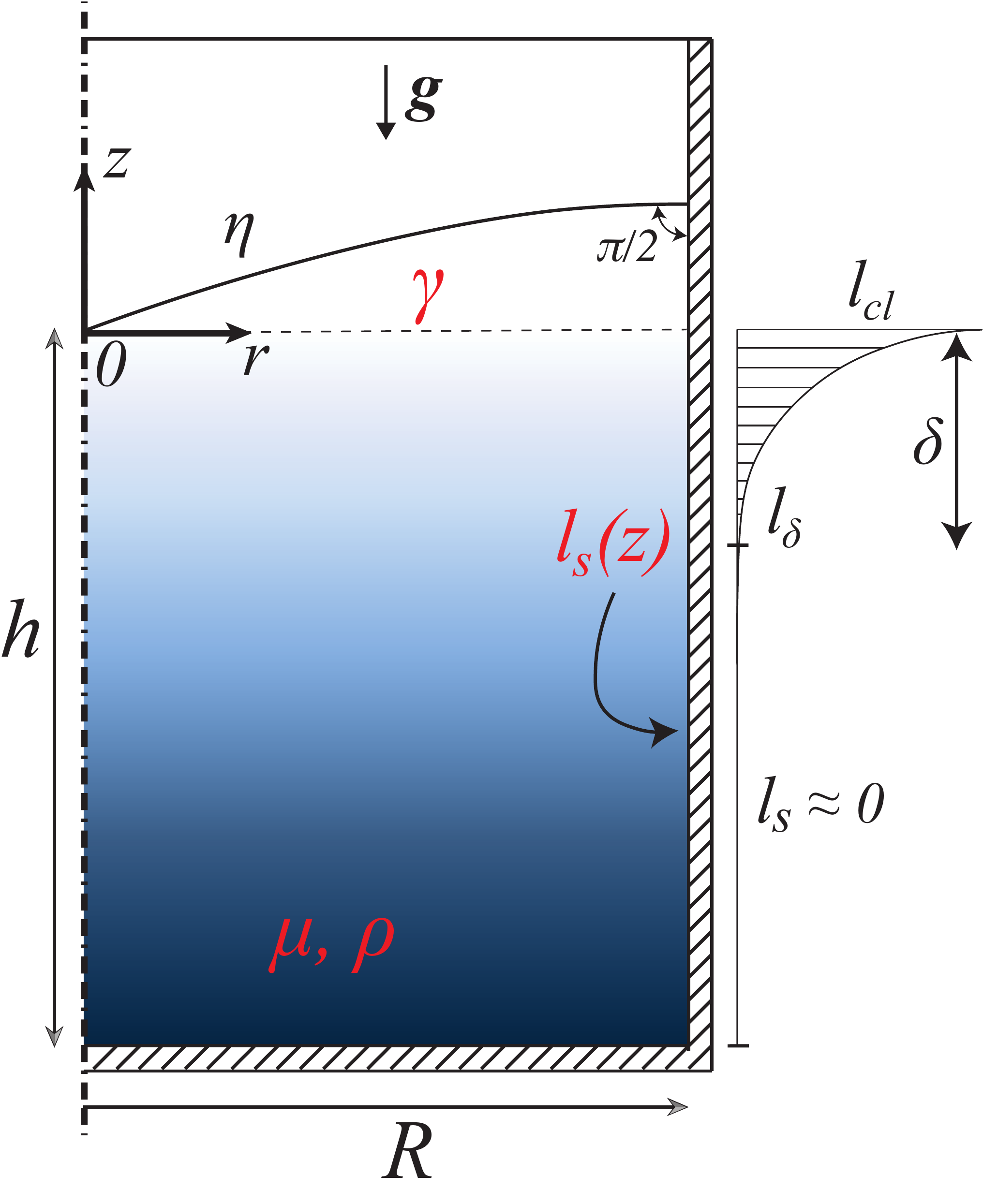}
\caption{Sketch of a circular-cylindrical container of diameter $D=2R$ filled to a depth $h$ with a liquid of density $\rho$ and
kinematic viscosity $\nu=\mu/\rho$. The air-liquid surface tension is $\gamma$. The origin of the Cartesian coordinate system is fixed at the center of the free liquid surface at rest, while the bottom is placed at $z=-h$. The dashed-dotted line is the geometrical axis of symmetry. The interface elevation is denoted by $\eta$, whereas $g$ is the gravity acceleration. The macroscopic contact angle is assumed constant and equal to $\pi/2$ during the entire motion, i.e. the free surface intersect orthogonally the sidewall. $l_s\left(z\right)$ denotes the depth-varying slip-length imposed at the lateral wall.}
\label{fig:Fig1} 
\end{figure}

The natural dynamics of small amplitude viscous capillary-gravity waves within a cylindrical vessel is governed by the unsteady Stokes equations,
\begin{equation}
\label{eq:NS}
\nabla\cdot\mathbf{u}=0,\ \ \ \ \partial_t\mathbf{u}=-\nabla p+Re^{-1}\Delta\mathbf{u},
\end{equation}
\noindent which are obtained by linearizing the Navier-Stokes equation around the static configuration, $\mathbf{u}_{st}=\mathbf{0}$, $p_{st}=-z$, $\eta_{st}=0$, and which are made non-dimensional by using the container's characteristic length $R$ and the velocity $\sqrt{gR}$ (see Fig.~\ref{fig:Fig1}). Consequently, the Reynolds number is defined as $Re=\sqrt{g R^3}/\nu$. At the free surface, $z=0$, the linearized dynamic and kinematic boundary conditions hold,
\begin{subequations}
\begin{equation}
\label{eq:Dyn}
\left[\left(-p+\eta-Bo^{-1}\partial_{\eta}\kappa\left(\eta\right)\right)\mathbf{I}+Re^{-1}\left(\nabla\mathbf{u}+\nabla^T\mathbf{u}\right)\right]\cdot\mathbf{n}=\mathbf{0},
\end{equation}
\begin{equation}
\label{eq:Kin}
\partial_t\eta=u_{z},
\end{equation}
\end{subequations}
\noindent where $\partial_\eta\kappa\left(\eta\right)$ is the first order variation of the curvature associated with the small interface perturbation $\eta\left(r,\phi,t\right)$, 
\begin{equation}
\label{eq:LinCurv}
\partial_\eta\kappa\left(\eta\right)=\partial_{rr}\eta+r^{-1}\partial_r\eta+r^{-2}\partial_{\phi\phi}\eta,
\end{equation}
\noindent $\mathbf{n}=\left\{0,0,1\right\}^T$ is the unit vector normal to the interface, $\mathbf{u}\left(r,\phi,z,t\right)=\left\{u_r,u_{\phi},u_z\right\}^T$ is the velocity field and $p\left(r,\phi,z,t\right)$ is the pressure field. The azimuthal coordinate is $\phi$. The Bond number is defined as $Bo=\rho g R^2/\gamma$. At the bottom wall at $z=-h/R=-H$, the no-slip and no-penetration condition apply, so that $\mathbf{u}=\mathbf{0}$. At the contact line, $z=0$ and $r=1$, we impose the classic free-end edge condition, which prescribes a constant zero slope for the moving interface, 
\begin{equation}
\label{eq:FCL}
\partial_r\eta=0.
\end{equation}
\noindent At the lateral solid wall, $r=1$, it is legit to impose the no-penetration condition, i.e. $u_r=0$, however, as anticipated in Sec.~\ref{sec:Sec1}, the tangential velocity components require an \textit{ad}--\textit{hoc} treatment, which will be discussed in detail in Sec.~\ref{sec:Sec3}. In the following we proceed by assuming that the boundary conditions for $u_{\phi}$ and $u_z$ at $r=1$ are linear. Hence, the linear homogeneous system can be written in the following matrix compact form
\begin{equation}
\label{eq:MatrixAB}
\left(\mathcal{B}\partial_t-\mathcal{A}\right)\mathbf{q}=\mathbf{0},\ \ \text{with}\ \ 
\mathcal{B}=\left(\begin{matrix}
I & 0\\
0 & 0
\end{matrix}\right),\ \ 
\mathcal{A}=\left(\begin{matrix}
Re^{-1}\Delta & -\nabla\\
\nabla^T & 0
\end{matrix}\right).
\end{equation}
\noindent We note that the dynamic and kinematic boundary conditions~\eqref{eq:Dyn}-\eqref{eq:Kin} do not explicitly appear in~\eqref{eq:MatrixAB}, but they are imposed as conditions at the interface \cite{bongarzone2021relaxation,bongarzone2022amplitude}. More precisely, in the numerical scheme, the kinematic condition governing the state variable $\eta$ is implemented as an additional equation dynamically coupled with $\mathbf{u}$ and $p$ in~\eqref{eq:MatrixAB} (see Appendix B of Ref.~\onlinecite{bongarzone2021relaxation}), whereas the three stress components of the dynamic condition are enforced as standard boundary conditions in the corresponding components of the vectorial momentum equation. The solution can be then expanded in terms of normal modes in time and in the azimuthal direction as
\begin{subequations}
\begin{equation}
\label{eq:NMEup}
\mathbf{q}_{mn}\left(r,\phi,z,t\right)=\left\{\mathbf{u},p\right\}^T=\hat{\mathbf{q}}_{mn}\left(r,z\right)e^{\lambda_{mn} t+\text{i}m\phi}+c.c.\,,
\end{equation}
\begin{equation}
\label{eq:NMEeta}
\eta_{mn}\left(r,\phi,t\right)=\hat{\eta}_{mn}\left(r\right)e^{\lambda_{mn} t+\text{i}m\phi}+c.c.\,.
\end{equation}
\end{subequations}
\noindent Substituting the normal form~\eqref{eq:NMEup}-\eqref{eq:NMEeta} in system~\eqref{eq:MatrixAB}, we obtain a generalized linear eigenvalue problem,
\begin{equation}
\label{eq:GenEig}
\left(\lambda_{mn}\mathcal{B}-\mathcal{A}_m\right)\hat{q}_{mn}=\mathbf{0},
\end{equation}
\noindent with $\left(\hat{\mathbf{q}}_{mn}\left(r,z\right),\hat{\eta}\left(r\right)\right)=\left\{\hat{u}_r,\hat{u}_{\phi},\hat{u}_z,\hat{p},\hat{\eta}\right\}^T$ the full eigenmode and $\lambda_{mn}=-\sigma_{mn}+\text{i}\,\omega_{mn}$ the associated eigenvalue. $\sigma_{mn}$ and $\omega_{mn}$ denote, respectively, the corresponding damping coefficient and natural frequency. Here the indices $\left(m,n\right)$ represent the number of nodal circles and nodal diameters, respectively, with $m$ also commonly known as azimuthal wavenumber. Owing to the normal mode expansion~\eqref{eq:NMEup}-\eqref{eq:NMEeta}, we note that the operator $\mathcal{A}_m$ is complex and it depends on the azimuthal wavenumber, as $\phi$--derivatives produce $\text{i}\,m$ terms. Lastly, in order to regularize the problem at the container axis, $r=0$, depending on the selected azimuthal wavenumber $m$, different regularity conditions must be imposed \cite{liu2012nonmodal,Viola2018b}:
\begin{subequations}
\begin{equation}
\label{eq:AXm0}
|m|=0\,:\ \ \ \hat{u}_r=\hat{u}_{\phi}=\partial_r\hat{u}_z=\partial_r\hat{p}=0,
\end{equation}
\begin{equation}
\label{eq:AXm1}
|m|=1\,:\ \ \ \partial_r\hat{u}_r=\partial_r\hat{u}_{\phi}=\hat{u}_z=\hat{p}=0,
\end{equation}
\begin{equation}
\label{eq:AXm234}
|m|>1\,:\ \ \ \hat{u}_r=\hat{u}_{\phi}=\hat{u}_z=\hat{p}=0.\ \ \ \ \ \ 
\end{equation}
\end{subequations}

\section{\label{sec:Sec3}A modified depth-varying slip-length model}

 In the spirit of Ref.~\onlinecite{ting1995boundary}, we propose here the following phenomenological macroscopic and depth-dependent slip-length model
\begin{equation}
\label{eq:SLrphi}
u_{\phi}+l_s\left(z\right)\left(\partial_r\hat{u}_{\phi}-\hat{u}_{\phi}/r\right),\ \ \ u_z+l_s\left(z\right)\partial_r u_z=0,
\end{equation}
\noindent with $l_s\left(z\right)$ described by the exponential law
\begin{equation}
\label{eq:SLlaw}
l_s\left(z\right)=l_{cl}\,\exp{\left(-\left(z/\delta\right)\log{(l_{\delta}/l_{cl})}\right)},\ \ \ \ z\in\left[-H,0\right],
\end{equation}
\noindent where $l_{cl}$ is the slip-length value at the contact line, $r=1$ and $z=0$, while $l_{\delta}$ is its value at a distance $\delta$ below the contact line, $r=1$ and $z=-\delta$, with $\delta$ representing the size of the slip region \cite{ting1995boundary}. In principle, $l_{cl}$, $l_{\delta}$ and $\delta$ are all free parameters. However, keeping in mind that one aims at mimicking a stress-free condition in the vicinity of contact line and a no-slip condition after a certain distance $\delta$, the natural choice is $l_{cl}\gg1$ ($\sim 10^{2}$$\div$$10^{4}$) and $l_{\delta}\ll1$ ($\sim 10^{-4}$$\div$$10^{-6}$). The range of values proposed in brackets is based on the sensitivity analysis of Sec.~6 and on the discussion outlined in App.~\ref{sec:AppA}.\\
\indent The definition of the slip region penetration depth, $\delta$, is more challenging. Miles (1990) \cite{miles1990capillary} postulated that the slip-length $l_s$ is a function of the position along the vertical wall and vanishes at a distance comparable to the thickness of the Stokes boundary layer (expressed in non-dimensional terms),
\begin{equation}
\label{eq:SBL}
\delta_{St}=\sqrt{2/\left(\omega_{mn} Re\right)},
\end{equation}
\noindent which represents the characteristic viscous length scale. Thus, in absence of any further informations, a reasonable choice for $\delta$ appears to be $\delta=\delta_{St}$.\\
\indent As the oscillation frequency $\omega_{mn}$ is not known \textit{a priori}, the major problem in imposing a variable slip-length depending on $\delta_{St}$ and, therefore, on $\omega_{mn}$, is that it turns the generalized eigenvalue problem~\eqref{eq:GenEig} in a nonlinear eigenvalue problem in $\lambda_{mn}=-\sigma_{mn}+\text{i}\,\omega_{mn}$, whose solution would required a fixed-point iteration-like scheme. Nevertheless, if the Reynolds number is sufficiently large,  i.e. $Re\gg 1$, the viscous correction to $\omega_{mn}$ is essentially negligible in norm, i.e. $\sigma_{mn}\ll\omega_{mn}$, so that one can simply assume $\omega_{mn}$ in~\eqref{eq:SBL} to be equal to the classical value derived in the potential flow limit and satisfying the well-known dispersion relation for inviscid capillary-gravity waves \cite{Lamb32}
\begin{equation}
\label{eq:DR}
\omega_{mn}^{inv}=\sqrt{\left(k_{mn}+k_{mn}^3/Bo\right)\tanh{\left(k_{mn}H\right)}}.
\end{equation}
\noindent In~\eqref{eq:DR} the wavenumber $k_{mn}$ is given by the $n$th--root of the first derivative of the $m$th--order Bessel function of the first kind satisfying $J_m'\left(k_{mn}\right)=0$. This implies that, as a first order approximation, one can define $\delta=\delta_{St}\approx\sqrt{2/\left(\omega_{mn}^{inv}Re\right)}$, hence making problem~\eqref{eq:GenEig} fully linear in $\lambda_{mn}$.\\
\indent We note that the value of Reynolds number explored in this work (see Sec.~\ref{sec:Sec5}) ranges from $Re\approx 1\,000$ to $100\,000$ ($\gg 1$), thus legitimizing the aforementioned approximation.

\section{\label{sec:Sec4}Numerical method}

The numerical scheme used in the eigenvalue calculation is a staggered Chebyshev--Chebyshev collocation method implemented in Matlab. The three velocity components are discretized using a Gauss--Lobatto--Chebyshev (GLC) grid, whereas the pressure is staggered on Gauss--Chebyshev (GC) grid. Accordingly, the momentum equation is collocated at the GLC nodes and the pressure is interpolated from the GC to the GLC grid, while the continuity equation is collocated at the GC nodes and the velocity components are interpolated from the GLC to the GC grid. This results in the classical $P_N$-$P_{N-2}$ formulation, which automatically suppress spurious pressure modes in the discretized problem. A two-dimensional mapping is then used to map the computational space onto the physical space. Lastly, the partial derivatives in the computational space are mapped onto the derivatives in the physical space, which depend on the mapping function. For other details see Refs.~\onlinecite{Viola2018b}, \onlinecite{Viola2018a}.
\indent The smallest length scale in the flow is represented by the thickness of the oscillating Stokes boundary layer, $\delta_{St}=\sqrt{2/\left(\omega_{mn}^{inv} Re\right)}$. In order to numerically solve the macroscopic flow at all scales, the computational grid used to discretized the eigenvalue problem~\eqref{eq:GenEig} must be smaller than such a length scale. The computationally most challenging case analyzed in this work (see Sec.~\ref{sec:Sec5}) corresponds to $H=h/R=3$, $Re=99\,045$ and $\omega_{mn}=10$ (non-dimensional frequency of mode $\left(m,n\right)=\left(3,10\right)$). For this case one has $\delta_{St}=0.0016$. By using a grid made of $100$ and $150$ radial and axial GLC nodes, respectively, the size of the first radial grid element at the lateral wall is $\Delta r_1=0.1571\,\delta_{St}$, whereas that of the first axial grid element at the free-surface (same for the bottom wall) is $\Delta z_1=0.2081\,\delta_{St}$. Smaller ratio $\Delta r_1/\delta_{St}$ and $\Delta z_1/\delta_{St}$ hold in all other cases examined. This ensures that the viscous boundary layers are properly accounted for and that results are sufficiently well converged.

\section{\label{sec:Sec5}Comparison with theoretical estimates}

\begin{figure*}[]
\includegraphics{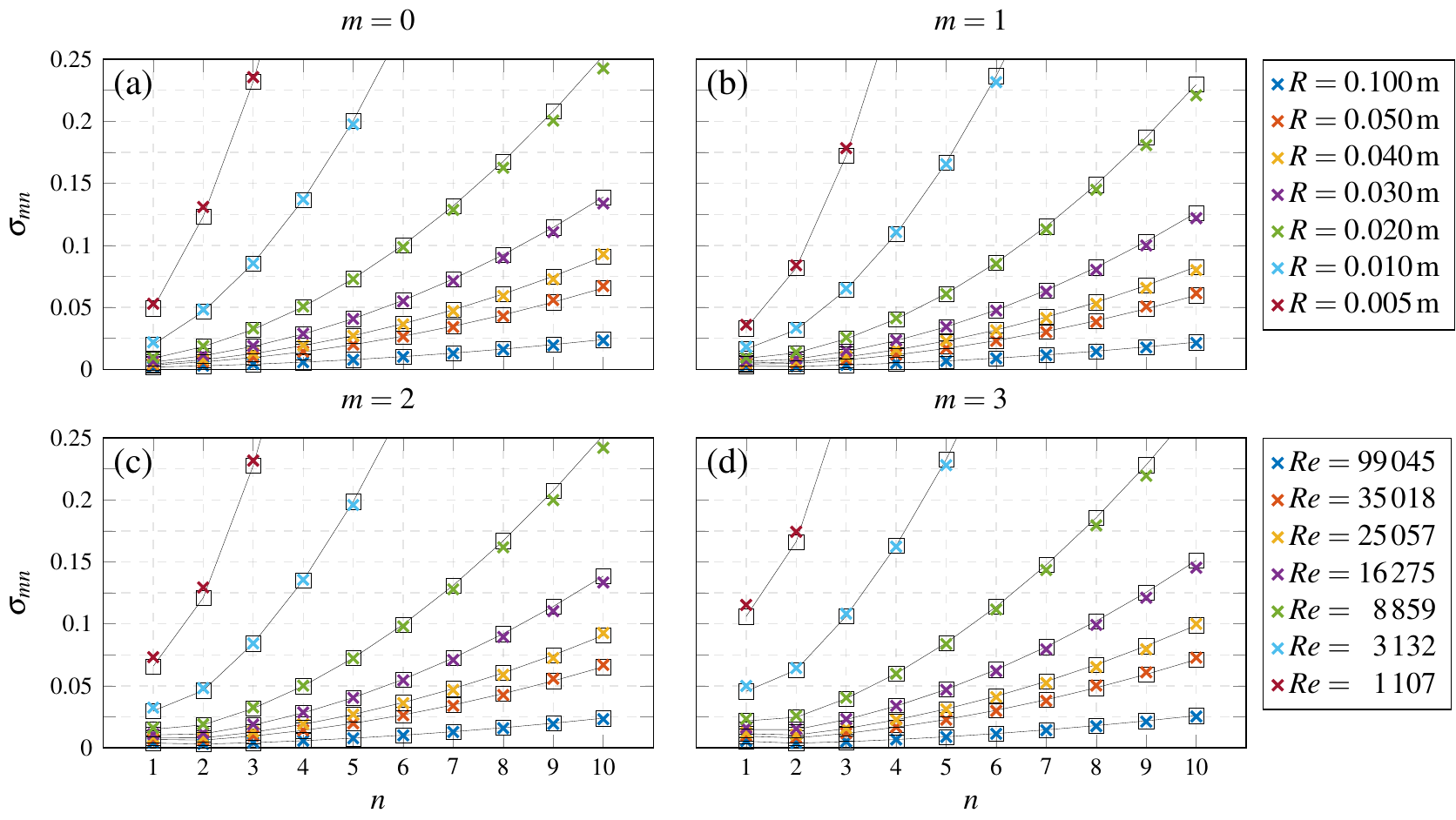}
\caption{Damping coefficients, $\sigma_{mn}$, of the first ten capillary-gravity waves for (a) $m=0$, (b) $m=1$, (c) $m=2$ and (d) $m=3$ and corresponding to seven different container radii, $R$ (see legend of panel (b)) and to a fluid depth $H=3$. The associated $Re=\sqrt{gR^3}/\nu$ is reported in the legend of panel (d). Black empty squares: theoretical prediction according to~\eqref{eq:DampCP}. Colored crosses: present numerical estimates. The black solid lines only serves to guide the eyes. The free slip-length model parameters are set to: $l_{cl}=10^3$, $l_{\delta}=10^{-5}$ and $\delta=\delta_{St}$.}
\label{fig:Fig2}
\end{figure*}

\begin{figure*}[]
\includegraphics{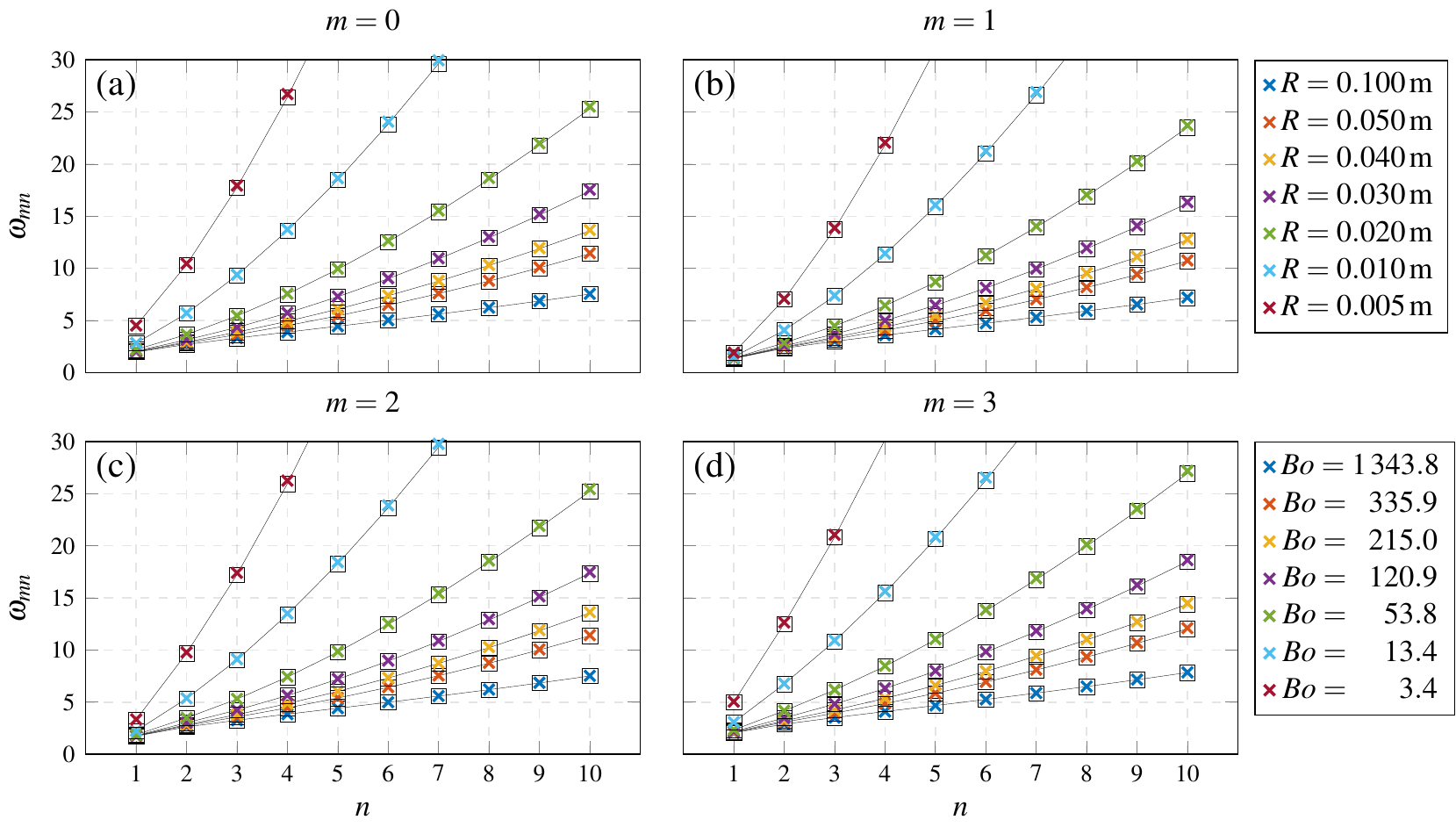}
\caption{Same as Fig.~\ref{fig:Fig2}, but in terms of natural frequencies, $\omega_{mn}$. The associated $Bo=\rho gR^2/\gamma$ is reported in the legend of panel (d). Black empty squares: theoretical prediction according to~\eqref{eq:FVC}. Colored crosses: present numerical estimates.}
\label{fig:Fig3}
\end{figure*}

\begin{figure}[t!]
\includegraphics[width=0.45\textwidth]{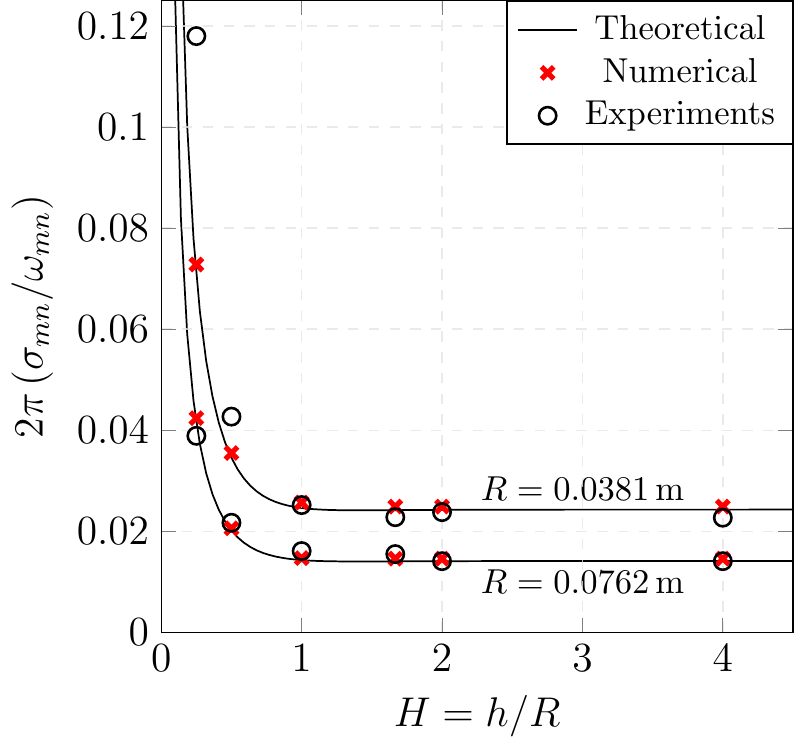}
\caption{Results for the damping coefficient versus non-dimensional fluid depth $H=h/R$ for the $\left(m,n\right)=\left(1,1\right)$ surface oscillations in a polished right circular cylinders of radius $R=0.0381\,\text{m}$ and $0.0762\,\text{m}$. Results are expressed in terms of logarithmic decrement $2\pi\sigma_{11}/\omega_{11}$. Black solid lines: theoretical prediction according to~\eqref{eq:DampCP}. Black circles: experimental measurements by Case \& Parkinson (1957) \cite{Case1957}. Black crosses: present numerical results with $l_{cl}=10^3$, $l_{\delta}=10^{-5}$ and $\delta=\delta_{St}$.}
\label{fig:Fig4}
\end{figure}

We neglect more complex phenomena acting at the contact line, e.g. hysteretic contact angle dynamics due to solid-like friction \cite{Keulegan59,Hocking87,voinov1976hydrodynamics,Dussan79,cox1986dynamics,Cocciaro91,Cocciaro93,jiang2004contact,dollet2020transition}, or at the free surface, e.g. surface contamination, etc.. Under these assumptions, capillary-gravity waves, over their motion, are damped by viscous dissipation occurring (i) at the oscillating Stokes layers at the solid bottom and sidewall, (ii) in the fluid bulk and (iii) at the free surface, the latter being typically negligible for ideal surface waves (clean surface). The damping coefficient associated with these three sources of dissipation was evaluated asymptotically in the limit of small kinematic viscosity $\nu$ (or large $Re\gg1$) by Case \& Parkinson (1957) \cite{Case1957} and by Miles (1967) \cite{Miles67}, who derived the well-known analytical approximation,
\begin{eqnarray}
\label{eq:DampCP}
\sigma_{mn}^{th}=\underbrace{\frac{2}{Re}k_{mn}^2}_{bulk}+\underbrace{\sqrt{\frac{\omega_{mn}^{inv}}{2Re}}\left(\frac{k_{mn}}{\sinh{\left(2k_{mn}H\right)}}\right)}_{bottom}+\\
+\underbrace{\sqrt{\frac{\omega_{mn}^{inv}}{2Re}}\left[\frac{1}{2}\,\frac{1+\left(m/k_{mn}\right)^2}{1-\left(m/k_{mn}\right)^2}-\frac{k_{mn}H}{\sinh{2k_{mn}H}}\right]}_{sidewall},\notag
\end{eqnarray}
\noindent with the superscript $^{th}$ standing for \textit{theoretical}. Based on~\eqref{eq:DampCP}, Henderson \& Miles (1990) \cite{henderson1990single} proposed the following viscous correction to the inviscid natural frequencies
\begin{equation}
\label{eq:FVC}
\omega_{mn}^{th}\approx\omega_{mn}^{inv}\left(1-\sigma_{mn}^{th}/\omega_{mn}^{inv}\right).
\end{equation}
\indent In this section, we compare our numerical estimates with these theoretical predictions. The free slip-length model parameters are here set to $l_{cl}=10^3$, $l_{\delta}=10^{-5}$ and $\delta=\delta_{St}$. A sensitivity analysis to variations of these parameters is outlined in Sec.~\ref{sec:Sec6}. For convenience, the fluid properties are fixed hereinafter, i.e. pure water with $\rho=1000\,\text{kg/m$^3$}$, $\gamma=0.073\,\text{N/m}$ and $\mu=0.001\,\text{kg/(m$\,$s)}$. By varying the container radius, e.g. $R=\left[0.005,0.01,0.02,0.03,0.04,0.05,0.1\right]\,\text{m}$, we span a wide range of possible combinations of Reynolds and Bond numbers, i.e. $1\,100\lesssim Re \lesssim100\,000$ and $3\lesssim Bo\lesssim1\,350$.\\
\indent In Fig.~\ref{fig:Fig2} the numerically computed damping coefficient, $\sigma_{mn}$, associated with the first ten modes for $m=0,1,2,3$ is compared with the corresponding theoretical prediction according to~\eqref{eq:DampCP}. The non-dimensional fluid depth was kept constant and equal to $H=3$ (deep water regime).\\
\indent The agreement between the two estimates is in general fairly good in the all range of $Re$, with a maximum deviation of approximately $12\,$\%. It is important to note that such a deviation does not represent an error, but only a difference between theoretical and numerical predictions. Indeed, neither the asymptotic formula~\eqref{eq:DampCP}, which assumes full no-slip along the wall till the contact line, nor the present numerical model, based on the spatially dependent slip-length model~\eqref{eq:SLlaw}, are accurate representations of the actual sidewall boundary condition. They only serve as a practical (theoretical or numerical) phenomenological modeling of the actual condition and their resulting prediction should be therefore as closest as possible. The agreement appears striking if compared with a deviation of over $100\,$\% observed in Ref.~\onlinecite{Viola2018b}, where a small and constant slip-length was numerically imposed.\\
\indent Fig.~\ref{fig:Fig3} shows a similar comparison, but in terms of natural frequencies, with the theoretical estimates computed using~\eqref{eq:FVC}. In this case, the maximum deviation is about $1\,$\% only, hence confirming that the viscous correction to the natural frequencies when $Re\gg1$, e.g. $Re\gtrsim1000$, is only mild and, essentially, negligible.\\
\indent Lastly, with focus on the first non-axisymmetric mode, $\left(m,n\right)=\left(1,1\right)$, in Tab.~\ref{tab:Tab1} we compare the theoretical and numerical damping coefficient for different container radii as the fluid depth is varied in the range $H=\left[0.5,1,3\right]$. 
\begin{table}
\caption{\label{tab:Tab1} Estimated damping coefficient, i.e. theoretical, $\sigma_{11}^{th}$, versus numerical, $\sigma_{11}$, associated with mode $\left(m,n\right)=\left(1,1\right)$ for three different fluid (water) depths, $H=h/R$, and seven different container radii, $R$. The corresponding Reynolds and Bond numbers are reported in legend of panel (b) of Figs.~\ref{fig:Fig2} and~\ref{fig:Fig3}, respectively. The slip-length parameters are: $l_{cl}=10^3$, $l_{\delta}=10^{-5}$ and $\delta=\delta_{St}=\sqrt{2/\left(\omega_{11}^{inv}Re\right)}$. The maximum difference between theoretical and numerical estimates is approximately $10$\% for $R=0.01\,\text{m}$ and $H=3$.}
\begin{tabular}{c|cc|cc|cc}
\hline
 & \multicolumn{2}{c|}{$H=0.5$} & \multicolumn{2}{c|}{$H=1$} &  \multicolumn{2}{c}{$H=3$}\\ \hline
$R$ $\left[\text{m}\right]$ & $\sigma_{11}^{th}$ & $\sigma_{11}$ & $\sigma_{11}^{th}$ & $\sigma_{11}$ & $\sigma_{11}^{th}$ & $\sigma_{11}$ \\ \hline
$0.100$ & 0.00301 & 0.00306 & 0.00244 & 0.00250 & 0.00247 & 0.00256\\
$0.050$ & 0.00516 & 0.00527 & 0.00419 & 0.00437 & 0.00425 & 0.00444\\
$0.040$ & 0.00614 & 0.00631 & 0.00501 & 0.00521 & 0.00507 & 0.00535\\
$0.030$ & 0.00772 & 0.00798 & 0.00631 & 0.00663 & 0.00639 & 0.00670\\
$0.020$ & 0.01076 & 0.01117 & 0.00882 & 0.00933 & 0.00893 & 0.00952\\
$0.010$ & 0.01967 & 0.02067 & 0.01628 & 0.01752 & 0.01646 & 0.01811\\
$0.005$ & 0.03926 & 0.04168 & 0.03285 & 0.03572 & 0.03318 & 0.03592\\ \hline
\end{tabular}
\end{table}
\noindent Once again, the agreement is fairly good for all combination of $R$ and $H$, with a maximum deviation of approximately $10\,$\%.\\
\indent A similar comparison is shown in Fig.~\ref{fig:Fig4}, where Fig.~2 of Ref.~\onlinecite{Case1957} has been reproduced. For a carefully polished sidewall, the agreement between experiments and theoretical predictions is rather satisfactory, thus suggesting that the analytical approximation~\eqref{eq:DampCP} is adequate under such controlled conditions. The present numerical predictions seem to agree fairly well with these estimates.

\section{\label{sec:Sec6}Sensitivity to the slip-length free parameters}

\begin{figure}
\includegraphics[width=0.504\textwidth]{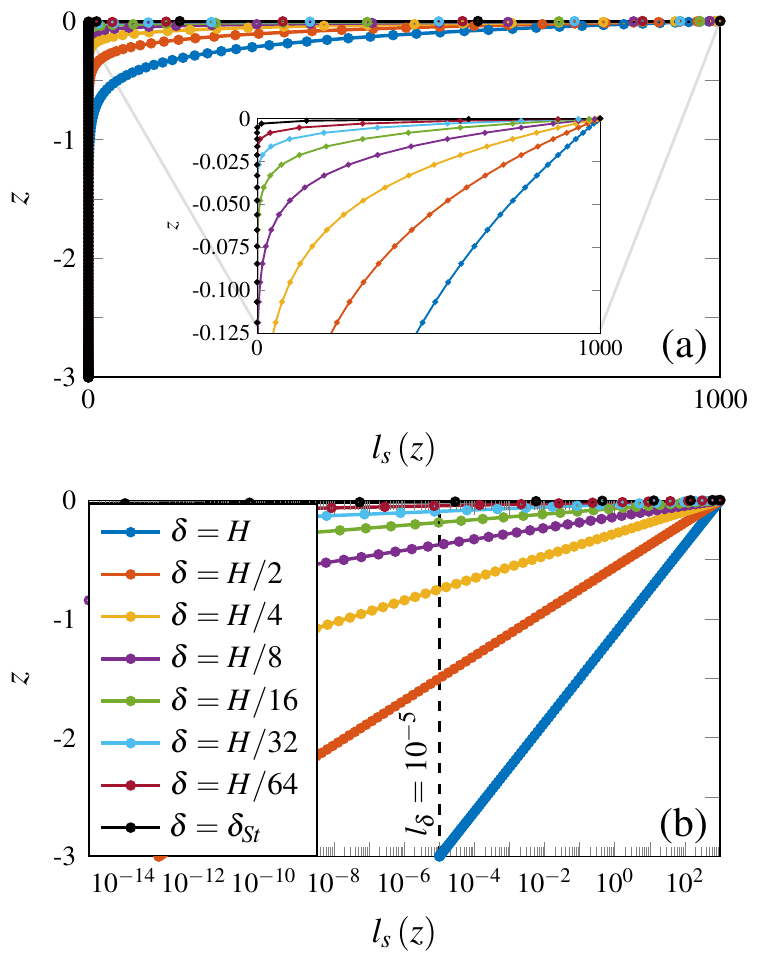}
\caption{Visualization of the exponential slip-length law, $l_s\left(z\right)$, along the sidewall, $z\in\left[-H,0\right]$, in (a) linear and (b) logarithm scale. The free parameters $l_{cl}$ and $l_{\delta}$ are set to $10^3$ and $10^{-5}$, respectively, whereas the size of the slip region $\delta$ is varied. The inset in (a) shows a zoom on the slip region. Container properties: $R=0.02\,\text{m}$ and $H=h/R=3$. The Stokes boundary layer $\delta_{St}=\sqrt{2/\left(\omega_{mn}Re\right)}$ used here corresponds to mode $\left(m,n\right)=\left(1,1,\right)$, for which $Re=8\,859$ and $\omega_{11}=1.3912$. Note that the colored filled circles corresponds to the actual grid points.}
\label{fig:FigA1}
\end{figure}

\begin{figure}
\includegraphics{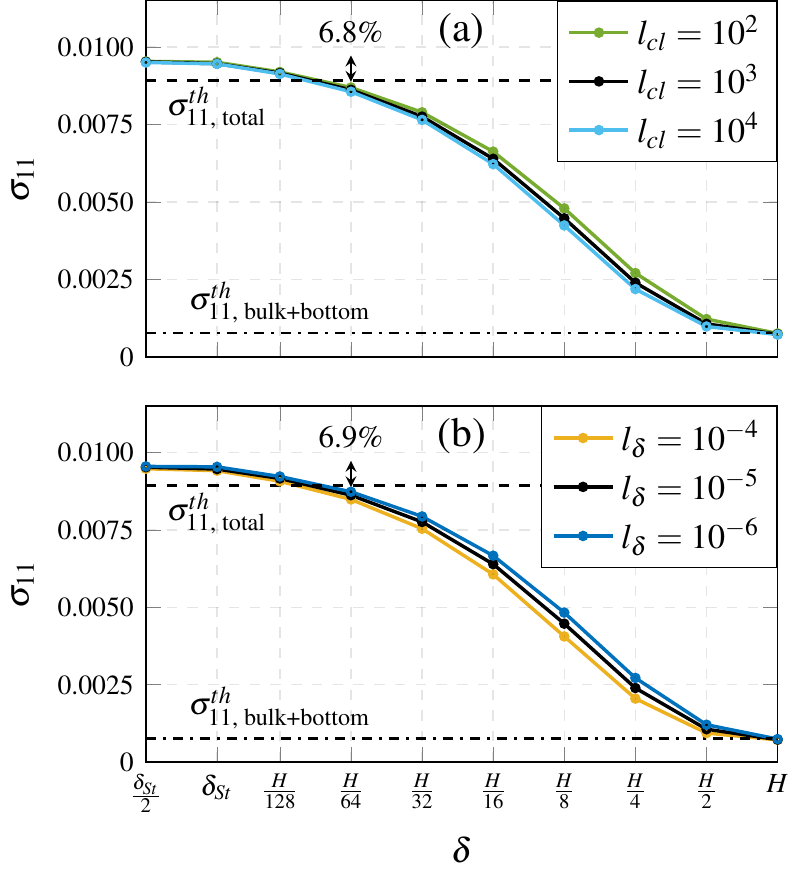}
\caption{Sensitivity analysis of the damping coefficient associated with the first non-axisymmetric mode, $\left(m,n\right)=\left(1,1\right)$, to the free parameters of the slip-length model. (a) Sensitivity to the value of $l_{cl}$ for varying $\delta$. (b) Sensitivity to the value of $l_{\delta}$ for varying $\delta$. Container properties: $R=0.02\,\text{m}$ and $H=h/R=3$. The black dash-dot line gives the theoretical estimate of the bulk and bottom contribution to the damping according to~\eqref{eq:DampCP}, while the black dashed line indicates the total value of $\sigma_{11}^{th}$ predicted by~\eqref{eq:DampCP}. }
\label{fig:FigA2}
\end{figure}

\begin{figure}[t!]
\includegraphics[width=0.485\textwidth]{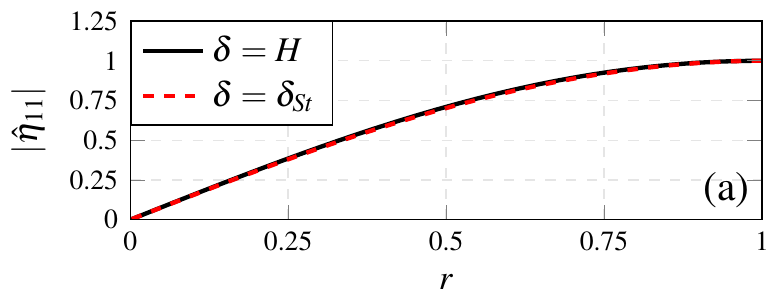}\\
\includegraphics[width=0.485\textwidth]{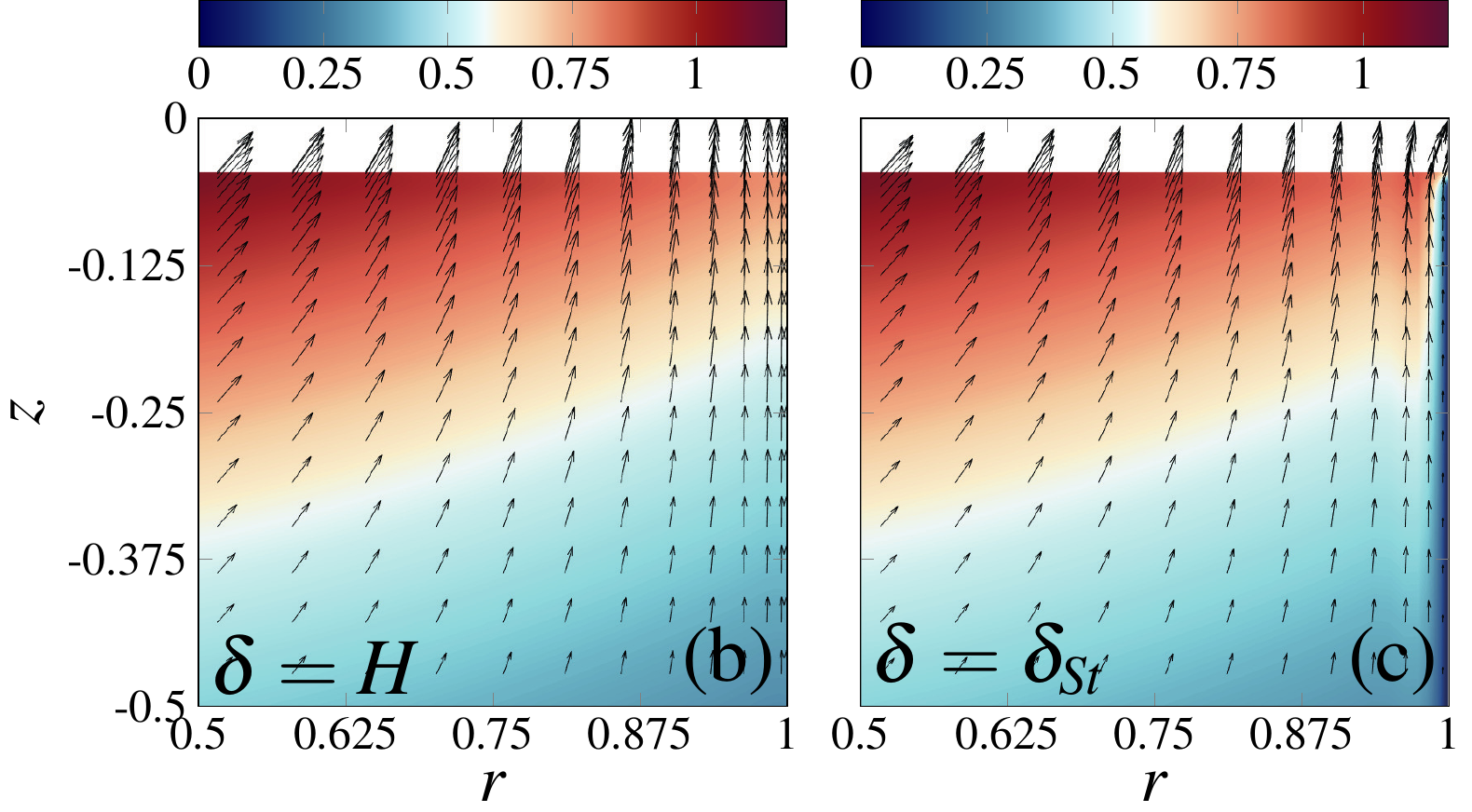}
\caption{(a) Absolute value of the eigen-interface associated with the first non-axisymmetric mode $\left(m,n\right)=\left(1,1\right)$, $\hat{\eta}_{11}\left(r\right)$ and computed for the two $\delta$ values, $\delta=H$ (black solid line) and $\delta_{St}$ (red dashed line). Container properties: $R=0.02\,\text{m}$ and $H=h/R=3$. The values $l_{cl}=10^3$ and $l_{\delta}=10^{-5}$ were used. No significant differences are observed between the two cases. Eigenmodes have been independently normalized by using the value of the interface at the contact line, so that $|\hat{\eta}_{11}\left(r=1\right)|=1$. (b)-(c) Absolute value of the azimuthal velocity component $\hat{u}_{\phi}\left(r,z\right)$ (filled 2D contour plot) and in-plane velocity components $\left\{\hat{u}_{r}\left(r,z\right),\hat{u}_{z}\left(r,z\right)\right\}$ (black arrows). The fields in (b) are computed using $l_{cl}=10^3$, $l_{\delta}=10^{-5}$ and $\delta=H$, whereas for the fields in (c) the same values of $l_{cl}$ and $l_{\delta}$ were used, but $\delta=\delta_{St}$.}
\label{fig:FigA3}
\end{figure}

As the slip-length model~\eqref{eq:SLlaw} requires the selection of three user-defined parameter values, i.e. $l_{cl}$, $l_{\delta}$ and $\delta$, in this section we explore the sensitivity of the damping coefficient, $\sigma_{mn}$, to variations of these parameters. Without loss of generality, we consider here, as a test case, the first non-axisymmetric eigenmode, $\left(m,n\right)=\left(1,1\right)$, for pure water and in a container of radius $R=0.02$ and $H=h/R=3$, for which $Re=8\,859$ and $Bo=53.8$. Similar considerations qualitatively were found to apply to all modes considered in Sec.~\ref{sec:Sec5}.\\
\indent A visualization of the slip-length variation along the sidewall, as the size of the penetration region $\delta$ is varied, is provided in Fig.~\ref{fig:FigA1}, where colored filled circles correspond to actual grid points in the numerical implementation.\\
\indent In Fig.~\ref{fig:FigA2}(a), the length of the slip region is varied in the range $\delta\in\left[\delta_{St}/2,H\right]$. The value of $l_{\delta}$ is fixed to $10^{-5}$, while three large values of $l_{cl}$, spanning three orders of magnitude, are considered. A similar analysis is displayed in Fig.~\ref{fig:FigA2}(b), where the value of $l_{cl}=10^3$ is kept fixed and three values of $l_{\delta}$ are used.\\
\indent As anticipated in Sec.~\ref{sec:Sec5}, both panels in Fig.~\ref{fig:FigA2} show that, as soon as sufficiently large and small end-values, respectively, of $l_s\left(z\right)$ are imposed at the contact line and at a distance $\delta$, the damping coefficient calculation is not very sensitive to variation of $l_{cl}$ and $l_{\delta}$. The range of values of $l_{cl}$ and $l_{\delta}$ used in this analysis is based on the reasoning described in App.~\ref{sec:AppA}. The only important parameter appears to be the size of the slip region, $\delta$. In Fig.~\ref{fig:FigA3} we show how a variation of $\delta$ between two cases with, respectively, $\delta=H$ and $\delta=\delta_{St}$ affects the velocity fields close to the wall and in the vicinity of the contact line region. In particular, one can see that when $\delta=H$, the entire lateral wall behaves as stress-free (see Fig.~\ref{fig:FigA3}(b)). On the contrary, when $\delta=\delta_{St}$, the sidewall is essentially a no-slip wall and, indeed, both the tangential velocity components go to zero in the Stokes boundary layer region as $r\rightarrow1$ (see Fig.~\ref{fig:FigA3}(c)), except for a very small region in the neighborhood of the contact line, i.e. $r\rightarrow1$ and $z\rightarrow0$.\\
\indent These considerations are consistent with the damping coefficient trend displayed in Fig.~\ref{fig:FigA2}. When $\delta=H$, the sidewall contribution to the damping coefficient is negligible, so that the total damping is generated in the bulk and in the solid bottom only (the latter is actually negligible in the deep water regime). By gradually reducing $\delta$, the damping increases and eventually, for $\delta\approx\delta_{St}$, it seems to saturate to a nearly constant value which is relatively close to the one predicted by~\eqref{eq:DampCP}, with a deviation of approximately $7$\% for the present case (see also Sec.~\ref{sec:Sec5}). With regard to Fig.~\ref{fig:FigA2}, we note that the typical penetration depth of mode $\left(1,1\right)$ is $2\pi/\left(Hk_{11}\right)=1.1375\approx1$, so that a variation of $\delta$ expressed as a fraction of the fluid depth $H$ is approximately equivalent to express the same variation as a fraction of the penetration length of the wave.

\section{\label{sec:Sec7}Conclusion}

In this manuscript, a phenomenological macroscopic and depth-dependent slip-length model for practical numerical implementation was presented. This model, inspired by that proposed by Ting \& Perlin (1995)\cite{ting1995boundary}, consists of a rapid exponential variation of the slip-length value along the sidewall, hence enabling one to simultaneously regularize the stress-singularity at the contact line and to properly account for the viscous boundary layer along the wall outside the slip contact line region.\\
\indent A thorough comparison with previous approximated theoretical estimates and a sensitivity analysis to the free model parameters was outlined, showing that the present model is suitable to predict with fair accuracy the damping coefficients and natural frequencies of confined viscous and small-amplitude capillary-gravity waves under controlled conditions, i.e. in absence of free surface contamination and complex contact angle dynamics. Although these conditions are only ideal and rarely met in real-life lab-scale experiments, the present slip-length boundary condition represents a functional sidewall condition for numerical implementation, which could allow one to revisit a wide range of existing experimental and numerically-based studies dealing with a moving contact line \cite{Viola2018b,bongarzone2021relaxation,bongarzone2022amplitude}. A straightforward generalization of the present model, would be the combination of the slip-length condition with the linear contact angle law proposed by Hocking (1987) \cite{Hocking87}. For instance, when these two conditions are combined, the artificial arbitrariness on the value of $l_{cl}$ in Eq.~\eqref{eq:SLlaw} is removed (see Refs.~\onlinecite{miles1990capillary} and~\onlinecite{ting1995boundary}) and rather replaced by a value in consistence with the proportionality constant linking the contact angle deviation to the interface velocity.\\
\indent Furthermore, the adoption of such a slip-length model offers room for further modeling, which could allow one to generalize it to more involved scenarios including, e.g., presence of a static meniscus, non-ideal wall wetting conditions, etc..\\
\indent Lastly, different depth-dependent slip-length laws could be tested, e.g. a gaussian law, a raised cosine distribution \cite{lacis2020steady} or other similar bell-like functions which assume a maximum value centered in $z=0$ (contact line) and quickly decaying. Some of these directions are being pursued and will be reported elsewhere.


\section*{Supplementary material}
See the supplementary material for further details on the derivation of the corrective factor $\beta$ discussed in App.~\ref{sec:AppA}.
\section*{Acknowledgments}
This research was supported by the Swiss National Science Foundation under grant 200021\_178971.\\
\indent The authors wish to thank Bastien Ravot the fruitful discussions on the modified Stokes second problem.
\section*{AUTHOR DECLARATIONS}
\subsection*{Conflict of interest}
The author have no conflicts to disclose.
\section*{Data Availability Statement}
The data that support the findings of this study are available from the corresponding author upon reasonable request.

\appendix
\section{\label{sec:AppA}Generalization of the theoretical estimates by accounting for a constant wall slip-length $l_s$}

Results presented in Sec.~\ref{sec:Sec5} and~\ref{sec:Sec6} have been computed by imposing a large $l_{cl}$ value, i.e. $\sim 10^{2}$$\div$$10^{4}$, and a small $l_{\delta}$ value, i.e. $\sim 10^{-4}$$\div$$10^{-6}$, in order to mimic, respectively, a stress-free condition in a small slip region in the neighborhood of the contact line and a no-slip condition outside that region along the wall. These two ranges of values have been estimated as discussed in the following.\\
\indent \, By considering a modified version of the Stokes second problem (in dimensional terms),
\begin{equation}
\label{eq:StokesII}
\partial_tu=\nu\partial_{zz}u,\ \ \ u\left(z\rightarrow\infty,t\right)=0,
\end{equation}
\noindent where the no-slip condition is replaced by the following slip-length condition
\begin{equation}
\label{eq:mod_slip}
u\left(z=0,t\right)-U\cos{\left(\omega t\right)}=l_s\partial_z u\left(z=0,t\right),
\end{equation}
\noindent with $U$ and $\omega$ the velocity and the frequency, respectively, of the oscillating moving plate, we can compute a corrective factor (due to the slip-length condition), $0\le\beta\le1$, to the classic dissipation induced by the viscous forces on the plate.\\ \indent This coefficient reads,
\begin{equation}
\label{eq:beta}
\beta=\frac{1+2\xi}{1+2\xi\left(1+\xi\right)},\ \ \ \ \xi=\frac{l_s}{\delta_{St}}=l_s\sqrt{\omega/\left(2\nu\right)}\ge0.
\end{equation}
\begin{figure}
\includegraphics[width=0.45\textwidth]{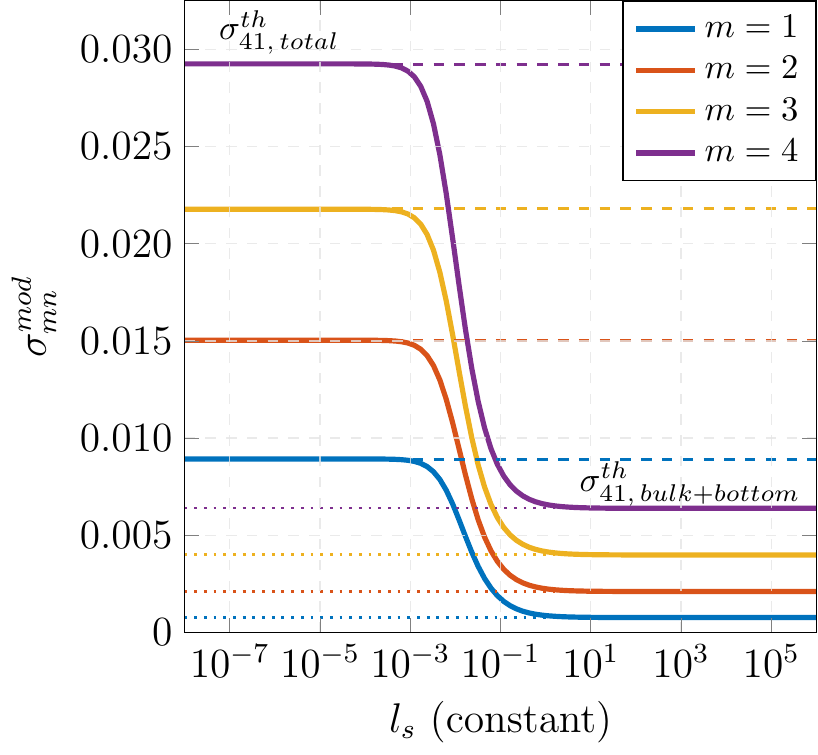}
\caption{Modified theoretical estimates of the damping coefficient of the first mode, $n=1$, associated with the five azimuthal wavenumbers $m=1,2,3,4$. The slip-length $l_s$ is constant. Curves were computed according to~\eqref{eq:DampCP} and accounting for the sidewall correction~\eqref{eq:DampCPslip}. Colored dotted lines correspond to $\sigma_{mn\text{, bulk+bottom}}^{th}$ (vanishing sidewall contribution), whereas colored dashed lines correspond to $\sigma_{mn\text{, total}}^{th}$. Parameters: $R=0.02\,\text{m}$, $H=3$ and pure water.}
\label{fig:FigA4}
\end{figure}
\noindent See supplementary material for its full derivation.\\
\indent Hence, the theoretical estimate by Case \& Parkinson (1957) \cite{Case1957} for the sidewall contribution to the damping coefficient could be generalized to the case of a constant slip-length ($l_s$) wall condition by accounting for such a correction as
\begin{equation}
\label{eq:DampCPslip}
\sigma_{mn\text{, sidewall}}^{mod}=\beta\sigma_{mn\text{, sidewall}}^{th},
\end{equation}
\noindent When $\xi\rightarrow0$ ($\beta=1$) and $\xi\rightarrow+\infty$ ($\beta=0$) the two limiting case of stress-free and no-slip sidewall boundary conditions, respectively, are retrieved.\\
\indent Fig.~\ref{fig:FigA4} shows the modified damping coefficient as the value of the constant slip-length is varied. For large $l_s$, i.e. $\gtrsim10^2$, the sidewall contribution to the damping is negligible and the wall behaves essentially as a stress-free wall. On the other hand, for small $l_s$, i.e. $\lesssim10^{-4}$, the original theoretical prediction by Case \& Parkinson (1957) \cite{Case1957} is retrieved, meaning that the sidewall behaves as a fully no-slip wall. The analysis outlined in Sec.~\ref{sec:Sec6} indeed confirmed that numerical results based on the depth-varying slip-length model~\eqref{eq:SLlaw} are not very sensitive to variations of $l_{cl}$ and $l_{\delta}$ values, once the beforehand mentioned limiting conditions are fulfilled.\\
\indent As a last comment, we note that the approach followed in the computation of $\beta$ (see supplementary notes) neglects the effect of the container curvature, by analogy with the asymptotic analysis by Case \& Parkinson (1957) \cite{Case1957}.






\section*{REFERENCES}
\bibliography{bibliography}
\end{document}